\documentclass{article}
\usepackage{spconf,amsmath,graphicx}
\usepackage{multirow}
\usepackage{amssymb, bm}
\usepackage{xcolor, tabu, cite}

\title{On end-to-end multi-channel time domain speech separation \\in reverberant environments}

\name{Jisi Zhang$^1$, Catalin Zorila$^2$, Rama Doddipatla$^2$ and Jon Barker$^1$
\thanks{\footnotesize \copyright 2020 IEEE ICASSP. DOI: 10.1109/ICASSP40776.2020.9053833}}

\address{$^1$University of Sheffield, Department of Computer Science, Sheffield, UK \\
        $^2$Toshiba Cambridge Research Laboratory, Cambridge, UK}

\begin{document}
\ninept
\maketitle
\begin{abstract}

This paper introduces a new method for multi-channel time domain speech separation in reverberant environments.
A fully-convolutional neural network structure has been used to directly separate speech from multiple microphone recordings, with no need of conventional spatial feature extraction.
To reduce the influence of reverberation on spatial feature extraction, a dereverberation pre-processing method has been applied to further improve the separation performance.
A spatialized version of wsj0-2mix dataset has been simulated to evaluate the proposed system.
Both source separation and speech recognition performance of the separated signals have been evaluated objectively.
Experiments show that the proposed fully-convolutional network improves the source separation metric and the word error rate (WER) by more than 13\% and 50\% relative, respectively, over a reference system with conventional features.
Applying dereverberation as pre-processing to the proposed system can further reduce the WER by 29\% relative using an acoustic model trained on clean and reverberated data.
\end{abstract}
\section{Introduction}
\label{sec:intro}
Speech separation is a challenging task that aims to segregate individual source speakers from a mixture signal. 
There are many applications that require performing source separation, such as speaker diarization, speaker verification or multi-talker automatic speech recognition in distant microphone scenarios. 
The recent advances in deep learning have facilitated the development of very powerful source separation systems by integrating neural networks and clustering algorithms~\cite{hershey2016deep,isik2016single,chen2017deep,kolbaek2017multitalker,wang2018supervised}.
Most previous methods operate on time-frequency domain signals, obtained by applying a Short Time Fourier Transform (STFT) to the time-domain signal, and mainly focus on separating signals received by a single microphone.
Recently, several end-to-end approaches have been proposed for the single channel speech separation task~\cite{luo2019conv,shi2019furcanext,Shi2019deep}, and have shown advantages over systems operating in the time-frequency domain.
The end-to-end approach extracts features from time-domain signals, jointly making use of magnitude and phase information, and outputs estimated time-domain signals, reducing the reconstruction errors caused by inaccurately estimated phase.

Microphone array processing has been demonstrated to be beneficial for speech enhancement and separation.
One approach is to combine fixed beamformers attending to a set of pre-defined angles with a single-channel separation network system~\cite{chen2017cracking,chen2018efficient}.
But the design of the beamformers requires the information of the microphone geometry and the discretised angles lead to a resolution problem.
An alternative approach is to combine spatial information and spectral information as input features to the separation network.
The inter-channel phase differences (IPDs) are common spatial features used as additional input for the separation network which has  
shown promising improvements for separation systems operating on both time and time-frequency domain signals ~\cite{wang2018multi,wang2018combining,chen2019multi,gu2019end,bahmaninezhad2019comprehensive}.
However, for methods applied to the time-domain signal, the window length of the convolutional kernel used for extracting spectral features is much smaller than the STFT window length used for extracting spatial features, causing a mismatch and misalignment problem.

In real environments, source separation and speech recognition, are made more challenging by the effects of reverberation (i.e., acoustic reflections).
Dereverberation algorithms such as the weighted prediction error (WPE) method~\cite{nakatani2010speech} are crucial for good performance. 
Systems for combined separation and dereverberation of a target speaker have been proposed in~\cite{delfarah2018recurrent,delfarah2019deep}.
A recurrent neural network for joint separation and dereverberation is trained to learn the clean target signal in the time-frequency domain from the reverberant mixture~\cite{delfarah2018recurrent}.
More recently, in~\cite{delfarah2019deep} an additional enhancement network was concatenated after the recurrent neural network for further enhancement of the separated signals.
However, both systems only use single-channel signals and are talker specific.
There remains a lack of study for joint separation and dereverberation in a multi-channel end-to-end fashion and for speaker-independent scenarios.

In this work we: (i) first design a system for time-domain speech separation using multiple microphones in a reverberant environment which solves the mismatch and misalignment problem mentioned above by employing 
a trainable 2-D convolutional layer to build a spatial encoder for spatial feature extraction from pairs of microphone signals , (ii) investigate the influence of reverberation on the separation task and find that applying dereverberation methods as a pre-processing stage can further improve the system's performance, and (iii) perform both speech separation and speech recognition experiments using a state-of-the-art acoustic model topology.

Speech separation methods have been applied in the past as a pre-processing step for multi-talker speech recognition and the results were promising for both single and multi-channel scenarios~\cite{isik2016single,bahmaninezhad2019comprehensive,menne2019analysis}. However, in previous studies, experiments have been conducted using either anechoic or reverberant environments, therefore is not straightforward to compare results between these conditions. In this work we present results in both anechoic and reverberant conditions. Additionally, we report results with dereverberation pre-processing for the latter case which, in the context of multi-channel time-domain end-to-end speech separation, is (to the best of our knowledge) a novel contribution.

The rest of the paper is organised as follows.
Section~\ref{sec:background} briefly describes the existing IPD-based end-to-end multi-channel separation approach. Section~\ref{sec:novel} then introduces our novel fully convolutional approach. Section~\ref{sec:experiment} presents implementation details and the experiment setup.
Results and analysis are presented in Section~\ref{sec:result}. Finally, the paper is concluded in Section~\ref{sec:conclusion}.

\section{Background}
\label{sec:background}

\begin{figure}[htp]
    \centering
    \includegraphics[width=0.48\textwidth]{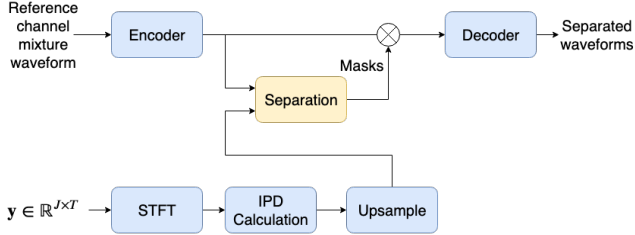}
    \caption{IPD based multi-channel speech separation approach.}
    \label{fig:ipd_tasnet}
\end{figure}

Recently, several end-to-end approaches have been proposed for the single channel speech separation task, one of which is the Conv-TasNet~\cite{luo2019conv}.
To exploit spatial features extracted from a microphone array, the Conv-TasNet has been extended to a multi-channel version in~\cite{gu2019end}, integrating both the separation network and IPD features, illustrated in the Figure~\ref{fig:ipd_tasnet}.

The Conv-TasNet takes the raw time-domain mixture as input to a neural network, which incorporates an encoder, a separator and a decoder, and estimates separate raw time-domain signals for each source.
Specifically, the encoder transforms each segment of the mixture time-domain to an $N$-dimensional representation, which is the input to the separation module for estimating masks. $N$ (typically 256) is the output dimension of the encoder.
The estimated masks will be multiplied with the mixture representation to generate separated representations for each source.
Then, the decoder reconstructs the estimated signals by inverting the separated representations back to time-domain signals.
Both the encoder and the decoder are 1-dimensional convolutional layers, and the separation block is a temporal fully-convolutional network (TCN)~\cite{lea2016temporal}.
The TCN is built from $R$ repetitions of a sub-block which stacks $X$ dilated 1-D convolutional blocks.
In each dilated 1-D convolutional block, the original convolution operation is replaced with a depthwise separable convolution~\cite{chollet2017xception} to reduce the number of parameters.

Conv-TasNet uses the utterance-level permutation invariant training (uPIT) criterion~\cite{kolbaek2017multitalker} and directly optimises the scale-invariant signal-to-noise ratio (SI-SNR) metric~(\ref{equ:si_snr}), also frequently used to assess the separation performance:
\begin{equation}
    \begin{split}
    & \text{SI\text{-}SNR} = 10\mathrm{log}_{10} \frac{||\text{s}_{target}||^2}{||\text{e}_{noise}||^2} \\
    & \text{s}_{target} = \frac{\big \langle \hat{s}, s \big \rangle s}{||s||^2},  \quad
    \text{e}_{noise} = \hat{s} - s_{target} 
    \end{split}
    \label{equ:si_snr}
\end{equation}
where $\hat{s}$ and $s$ denote the estimated and clean source, respectively, and $||s||^2=\big \langle s, s \big \rangle$ denotes the signal power.

The IPD feature represents the phase difference between a pair of signals in the time-frequency domain, calculated as
\begin{equation}
\label{equ:ipd}
    \mathrm{IPD_{nk}^{(ij)} = \angle Y_{nk}^{i} - \angle Y_{nk}^{j}},
\end{equation}
where $Y_{nk}^{i}$ denotes the STFT of the $i$-th microphone signal at frame $n$ and frequency bin $k$.

The IPD spatial features are upsampled to the same frame length as the single-channel spectral representation, concatenated with the spectral (single-channel) features and fed into the separation network.
\vspace{-10pt}
\section{Multi-channel end-to-end separation}
\label{sec:novel}

\begin{figure}[htp] 
\centering 
\includegraphics[width=0.48\textwidth]{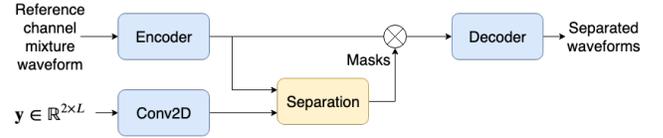} 
\caption{Proposed multi-channel speech separation approach.} 
\label{fig:multi_tasnet} 
\end{figure}

Instead of using conventional features such as IPDs, this work aims to design a novel separation model directly extracting spatial features from time-domain multi-channel signals with a 2-dimensional convolutional layer, keeping the whole system in an end-to-end framework, as illustrated in the Figure~\ref{fig:multi_tasnet}.
The 2-D convolutional layer has a kernel size of $(2, L)$, such that the analysis window has the same length $L$ as the 1-D convolutional layer for single channel encoder, thus keeping the number of frames of spatial features the same as the spectral representation.
The dimension of the output channel of the 2-D convolutional layer, $S$, which represents the spatial features, is fixed small, such that the kernel will focus more on spatial information across different channels.
A non-linear activation function, i.e., rectified linear unit (ReLU), follows the 2-D convolutional layer and together they construct the spatial encoder.
The spatial encoder takes a pair of signals as input and will be repeated to process all pairs when more than one pair is made.
Then, the spatial representations from each pair and spectral representations are concatenated along the channel dimension and fed into the subsequent separation block.

The spatial encoder is trained jointly with the main separation network.
The objective function for the multi-channel TasNet is $\text{SI\text{-}SNR}$, and the network is trained using uPIT.

When the number of microphone channels, $J$, is larger than 2, another option is to use a 2-D convolutional layer with a bigger kernel size, i.e. $(J, L)$, to take all the channels together as input.
Using this larger kernel and increasing the representation dimension achieve a similar separation performance to the pair-making system.
To make a fair comparison with the IPDs which is calculated from paired signals, the proposed system continues using the 2-D convolutional layer with the kernel size of $(2, L)$ in the following experiments.

\begin{figure}[htp]
    \centering
    \includegraphics[width=0.48\textwidth]{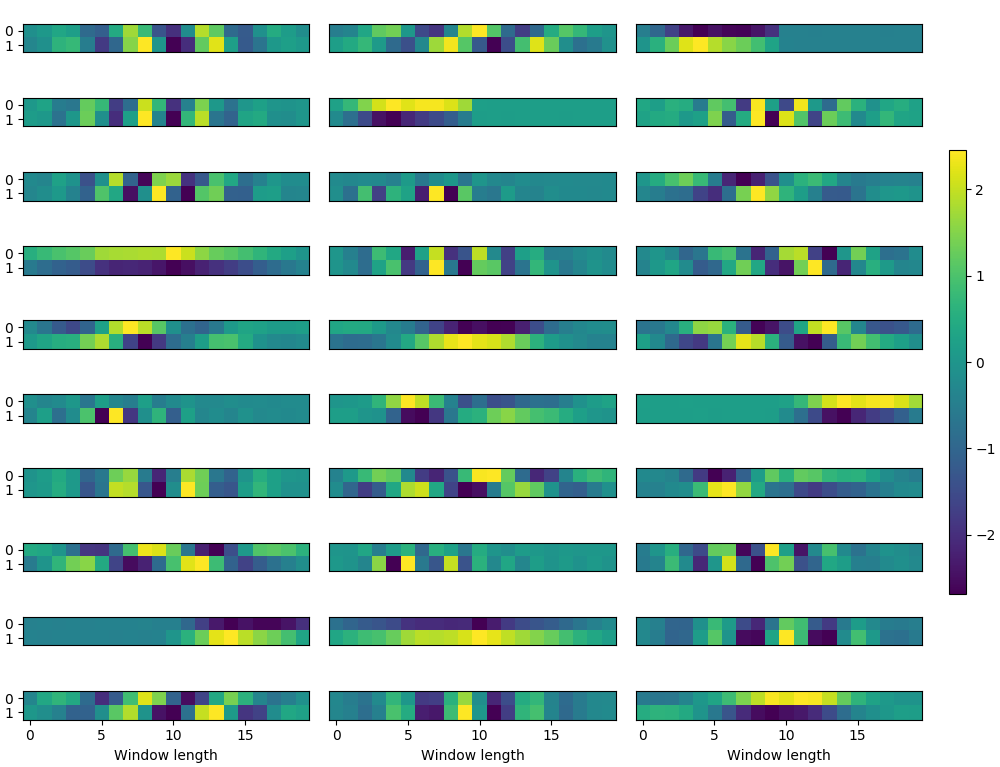}
    \caption{Basis functions of 2-D spatial encoder of proposed method.}
    \label{fig:spatial_basis}
		\vspace{-10pt}
\end{figure}

Figure~\ref{fig:spatial_basis} illustrates the learned filters of the 2-D convolutional layer.
The learned filters calculate the difference between two signals with different valid window lengths, which are all much shorter than the normal window length of the STFT, indicating the network itself can find better
filters to encode the spatial features.

The estimation of spatial features will decline in quality in a reverberant environment.
Therefore, we have investigated the effect of dereverberation on the multi-channel mixtures and unmixed targets before training and testing. The multiple input multiple output version of the Weighted Prediction Error (WPE) method was used for dereverberation~\cite{nakatani2010speech}.
WPE is selected since it can conserve the spatial differences at different microphone positions, which is required for subsequent microphone array processing~\cite{yoshioka2012generalization}.

After the dereverberation, the separation network is trained to separate the processed mixture signal to individual anechoic target signals.
To be precise, during the training, the input to the network is the processed mixture signals and the targets are the anechoic premixed signals.
During the testing, the multi-channel mixtures are pre-processed by the WPE before being fed into the separation network.

\vspace{-10pt}
\section{Experiment Setup}
\label{sec:experiment}
\vspace{-5pt}
\subsection{Data simulation}

Multi-channel reverberant data for training and evaluation has been simulated using the Wall Street Journal corpus, in a similar fashion to the spatialized WSJ-2mix~\cite{wang2018multi}.
Specifically, for each mixture utterance, a room configuration (length-width-height) is uniformly sampled from 5-5-3 m to 10-10-4 m.
In the centre of each room is located a circular array with 6 elements arranged uniformly with an angle of separation of 60 degrees.
The radius of the array is uniformly drawn from a range between 7.5 cm and 12.5 cm.
Two speakers are randomly positioned in the central area of the room.
Then the classic image method~\cite{allen1979image} has been used to generate multi-channel room impulse responses with $T_{60}$ uniformly distributed between 200 ms and 600 ms. 
The multi-channel signals are obtained by convolving the clean speech with the room impulse responses.
Two speakers' speech is mixed using the same utterance-pairings and the corresponding  signal-to-noise (SNR) values as in~\cite{wang2018multi}. Note, to match the exact SNR values in ~\cite{wang2018multi}, which range from aboute -2.5 dB to 2.5 dB, the anechoic and reverberant source images are rescaled.

The size of the training, validation, and test sets are 20k, 5k, and 3k utterances, respectively.
The sampling rate of all the data is 8 kHz.
The speakers of the utterances used for testing do not appear during training, i.e., the system is being evaluated with a \textit{speaker-independent} speech separation task.

Data with no reverberation, 
but with identical microphone and source geometry,
has also been simulated to show the separation performance in the anechoic environment and for use in the dereverberation experiments.

\subsection{Acoustic model}
 
To evaluate the speech recognition performance, two acoustic models have been trained using Wall Street Journal corpus. One model (AM1) was trained on roughly 80~hrs of clean WSJ-SI284 data, and the other one (AM2) was trained on the same clean data plus a spatialized version of it (in total roughly 160~hrs of training data). Randomly selected room impulse responses used to simulate the multi-channel mixtures above were employed to reverberate the train set of AM2.
The audio data is downsampled to 8 kHz to match the sampling rate of data used for separation experiments.
The acoustic model topology is a 12-layered Factorised TDNN~\cite{povey2018semi}, where each layer has 1024 units.
The input to the acoustic model is 40-dimensional MFCCs and a 100-dimensional i-Vector.
A 3-gram language model is used during recognition.
The acoustic model is implemented with the Kaldi speech recognition toolkit~\cite{povey2011kaldi}.
With our set-up the ASR results obtained with AM1 on the standard clean WSJ Dev93 and Eval92 are 7.2\% and 5.2\% WER, respectively.

\subsection{Separation network configuration}

The hyper-parameters of Conv-TasNet are set as those that produced best performance in the original paper \cite{luo2019conv} 
namely, $R=3$, $X=8$, $L=20$, and the batch size $M=3$.
The number of filters, $S$, in the 2-D convolutional layer is set to 30.
Each utterance is split into multiple segments with a fixed length of 4 seconds.
Pairs of signals, (1, 4), (2, 5), (3, 6), (1, 2), (3, 4) and (5, 6), are selected for all multi-channel experiments with 6 channels.
The experiments with 2 channels all use the pair (1, 4), on opposite sides of the circular array, such that the distance between the selected microphones equals the array's diameter.

The IPD system with fixed kernels in~\cite{gu2019end} has been reproduced as a baseline, with the window length of the STFT set to 32 samples.
During implementation, the 1-D convolution kernel to encode multi-channel information is initialized by the STFT parameters and then fixed during the training.
The spatial feature input to the separation block is $\mathrm{cos(IPD) + sin(IPD)}$.

Different variants of TasNet have been trained to attempt to solve the combined separation and dereverberation problem.
Both the single-channel system and the multi-channel system have been trained in an end-to-end fashion, with the input being the reverberant mixture signals and the training targets being the clean target signals.
These systems are provided to allow comparison with the system using the WPE as a preprocessing step, showing the problem of extracting spatial features directly from reverberant signals.

\vspace{-10pt}
\section{Results}
\label{sec:result}

\subsection{Speech separation performance}

Proposed and reference methods have been evaluated in terms of SI-SNR improvements in both anechoic and reverberant conditions (Table~\ref{tab:sisnr}). Although tested on different data sets, the baseline single-channel TasNet has achieved performance similar to that reported in~\cite{luo2019conv} for the anechoic condition. However, the performance drops noticeably (nearly 8dB) in the reverberant case, indicating that reverberation has a strong impact on the source separation performance of standard TasNet. The performance is partially recovered using the 
second reference system combining single-channel TasNet and the 6-channel IPD features which has yielded significant SI-SNR improvements in both conditions compared with the plain single-channel TasNet. The gains of the IPD system are consistent with those published in the literature~\cite{gu2019end} (i.e., 11.5 dB in reverberation), and the small differences are due to the slightly different data simulation setup (e.g., in this paper the radius of the circular array is randomly varied instead of being fixed).

Concerning the proposed speech separation method, by using 2 channels has either outperformed or matched the performance of the 6-channel IPD system in the anechoic and reverberant conditions, respectively. Using 6 channels led to almost perfect anechoic separation and yielded more than 13\% relative SI-SNR improvement over the reference IPD system in reverberation, suggesting 
that the separation algorithm can benefit from more spatial information extracted from different pairing signals.
The superiority of the proposed system over the one based on the IPD can be explained as follows.
The proposed spatial features extracted using 2-D convolutional layers are aligned with the learned spectral features from the encoder, whereas for IPD extraction the STFT window length is larger than the window length $L$ of the convolution kernel, therefore the IPD features need upsampling which may cause misalignment with the spectral features.

\begin{table}[t]
\centering
\caption{SI-SNR improvements of reference and proposed systems.}
\label{tab:sisnr}
\begin{tabular}{lccc}
\hline
\multirow{2}{*}{System} &\multirow{2}{*}{\#chs} & \multicolumn{2}{c}{SI-SNRi (dB)}                   \\ \cline{3-4} 
        &       & \multicolumn{1}{c}{Anechoic} & \multicolumn{1}{c}{Reverb} \\ \hline
TasNet~\cite{luo2019conv}  &1   & 14.6 & 6.7  \\ % old: 16.0 7.2 -> 14.6 6.7
Tasnet+IPD~\cite{gu2019end}&6   & 19.7 & 10.9 \\
Proposed                   &2   & 28.2 & 10.9 \\
Proposed                   &6   & \textbf{30.0} & \textbf{12.6} \\ \hline
\end{tabular}
\vspace{-10pt}
\end{table}

\vspace{-10pt}
\subsection{Speech recognition performance}
The results of the speech recognition experiments are depicted in Table~\ref{tab:asr_reverb}.
In the anechoic condition, the WER of the single-channel Tasnet is around 19\%, which is significantly better than the results based on GMM-HMM~\cite{isik2016single} or end-to-end ASR models~\cite{seki2018purely,chang2019end}.
Remarkably, all the multi-channel separation models are able to achieve a speech recognition performance close to the Oracle result. The AM with multi-condition training (AM2) achieves similar accuracy compared with the AM trained on clean, 
since the signals are well separated and there is no reverberation.

In the reverberant case, the proposed multi-channel separation system provides both acoustic models with consistent performance gains over the IPD system.
Remarkably, for AM2, the proposed system with 6 channels yields more than 50\% relative WER improvement over the IPD system with 6 channels.
An interesting observation is that although the SI-SNR improvement in Table~\ref{tab:sisnr} for the reference IPD system and the proposed 2-channel method in reverberation are similar, the the corresponding WER results in Table~\ref{tab:asr_reverb} are significantly different. 
This observation requires further analysis in the future, and it may show that the SI-SNR metric may not be optimal when deploying speech separation enhancement for multi-talker ASR.
AM2 performs much better than AM1, since AM2 is trained with multi-condition data, thus is more robust to the distortions introduced by the reverberation and the separation model.
However, the ASR gain from the separation pre-processing in the reverberant condition is still less than that in the anechoic condition, suggesting that the system has difficulties in extracting spectral features and spatial features from signals with reverberation. Therefore, next we investigated the effect of dereverberation and speech separation on the ASR accuracy (Table~\ref{tab:asr_wpe}). 

\begin{table}[t]
\centering
\caption{ASR accuracies in WER(\%) of reference and proposed systems for anechoic and reverb conditions.}
\label{tab:asr_reverb}
\begin{tabu}{lccccc}
\hline
\multirow{2}{*}{System} & \multirow{2}{*}{\#nchs} & \multicolumn{2}{c}{Anechoic} &\multicolumn{2}{c}{Reverb} \\ \cline{3-6} 
& & AM1    & AM2 & AM1    & AM2           \\ \hline
Mixture                     & 1     & 79.6  & 80.8  & 88.6  & 82.4   \\
TasNet~\cite{luo2019conv}    & 1     & 19.3  & 18.5 & 74.5  & 47.1      \\
Tasnet+IPD~\cite{gu2019end} & 6     & 10.6  & 10.3  & 70.0  & 40.3  \\
Proposed                    & 2     & 10.7  & 10.6  & 62.2  & 25.3  \\
Proposed                    & 6     & 9.2   & 9.2   & 57.2  & 19.8  \\
\rowfont{\color{gray}}
Oracle                      & 1     & 9.1   & 9.1   & 46.2  & 11.3 \\ \hline
\end{tabu}
\vspace{-5pt}
\end{table}

Three subsets of experiments were performed. Firstly, the reverberant unprocessed speech mixtures and the clean (anechoic) targets were used to train the separation systems, thus constraining the networks to learn both to separate the speakers and perform dereverberation. Secondly, both mixtures and targets were enhanced using WPE during training, and, lastly, the WPE mixtures and the anechoic targets were employed for training.

The results for the single-channel TasNet show that the system is able to jointly separate and dereverberate the signals (second row in Table~\ref{tab:asr_wpe}), but the ASR accuracy is lower than in the WPE-WPE case (third line). Interestingly, the performance in the third case (WPE-clean) is slightly worse for AM2 compared with the WPE-WPE case, indicating that the spectral representation alone cannot benefit from the cleaner targets during training.

Concerning the proposed 6-channel separation system, similar trends can be observed, but the absolute improvements are significantly larger. In this case, for the WPE-Clean condition there is a consistent WER improvement with both acoustic models compared with the WPE-WPE case, indicating that a more powerful spectral and spatial signal representation can benefit from cleaner targets during training, i.e., the network is able to remove the mild reverberation remained in the WPE processed data. The acoustic model trained with reverberant data (AM2) outperforms the model trained only with clean speech (AM1), however, applying dereverberation has reduced the performance gap between the models.

These results indicate that combining spectral and spatial signal representations in an end-to-end fashion helps improve speech separation and ASR accuracy. Also, dereverberation pre-processing can yield significant performance improvement.
Further research is ongoing to extend this system to a multi-device scenario and to evaluate the separation performance with real data recorded in realistic environments such as, for example, CHiME-5~\cite{barker2018fifth}.

\begin{table}[t]
\small
\centering
\caption{Results of different source separation and dereveberation strategies. Targets' enhancement is during training only.}
\label{tab:asr_wpe}
\begin{tabular}{lccccc}
\hline
\multirow{2}{*}{System}  & \multicolumn{2}{c}{Enhancement} & \multicolumn{2}{c}{WER(\%)} & \multirow{2}{*}{\shortstack{SI-SNRi\\(dB)}} \\ \cline{2-5}
    & Mixture & Targets & AM1 & AM2 &   \\ \hline
\multirow{4}{*}{\shortstack{TasNet~\cite{luo2019conv}\\(1-ch)}}& None & None  & 74.5 & 47.1 & 6.7       \\
& None        & Clean       & 53.3 & 44.1 & 8.7   \\
& WPE         & WPE         & 39.0 & 30.3 & 9.6    \\ 
& WPE         & Clean       & 38.4 & 31.8 & 8.8     \\ \hline
\multirow{4}{*}{\shortstack{Proposed\\(6-ch)}}& None       & None& 57.2 & 19.8 & 12.6     \\
& None        & Clean       & 25.2 & 20.0 & 15.2      \\
& WPE         & WPE         & 20.0 & 14.4 & 15.5     \\ 
& WPE         & Clean       & 16.3 & 14.0 & 16.0   \\ \hline
\end{tabular}
\vspace{-10pt}
\end{table}

\vspace{-10pt}
\section{Conclusions}
\label{sec:conclusion}
\vspace{-5pt}
In this paper, we argued that conventional spatial features are not optimal for an end-to-end time domain speech separation system.
Using a trainable kernel with a window length matched to that of the spectral encoder can efficiently address the misalignment and mismatch problem, leading to a better multi-channel separation performance.
A further investigation shows that, in realistic environments, reverberation will degrade the spatial feature extraction.
Applying dereverberation methods to the mixture signals as a preprocessing step, allows better spatial features to be extracted by the proposed multi-channel system, improving the speech separation performance in reverberant environments to a similar level to that achieved using a single-channel method in an anechoic environment.

\vfill\pagebreak

\bibliographystyle{IEEEbib}
\bibliography{refs}

\begin{thebibliography}{10}

\bibitem{hershey2016deep}
J.~R. Hershey, Z.~Chen, J.~Le~Roux, and S.~Watanabe,
\newblock ``Deep clustering: Discriminative embeddings for segmentation and
  separation,''
\newblock in {\em Proc. ICASSP}, 2016, pp. 31--35.

\bibitem{isik2016single}
Y.~Isik, J.~Le~Roux, Z.~Chen, S.~Watanabe, and J.~R. Hershey,
\newblock ``Single-channel multi-speaker separation using deep clustering,''
\newblock {\em Interspeech 2016}, pp. 545--549, 2016.

\bibitem{chen2017deep}
Z.~Chen, Y.~Luo, and N.~Mesgarani,
\newblock ``Deep attractor network for single-microphone speaker separation,''
\newblock in {\em Proc. ICASSP}, 2017, pp. 246--250.

\bibitem{kolbaek2017multitalker}
M.~Kolb{\ae}k, D.~Yu, Z.-H. Tan, J.~Jensen, M.~Kolbaek, D.~Yu, Z.-H. Tan, and
  J.~Jensen,
\newblock ``Multitalker speech separation with utterance-level permutation
  invariant training of deep recurrent neural networks,''
\newblock {\em IEEE/ACM Trans. Audio, Speech, Lang. Process.}, vol. 25, no. 10,
  pp. 1901--1913, 2017.

\bibitem{wang2018supervised}
D.~Wang and J.~Chen,
\newblock ``Supervised speech separation based on deep learning: An overview,''
\newblock {\em IEEE/ACM Trans. Audio, Speech, Lang. Process.}, vol. 26, no. 10,
  pp. 1702--1726, 2018.

\bibitem{luo2019conv}
Y.~Luo and N.~Mesgarani,
\newblock ``{Conv-TasNet}: Surpassing ideal time--frequency magnitude masking
  for speech separation,''
\newblock {\em IEEE/ACM Trans. Audio, Speech, Lang. Process.}, vol. 27, no. 8,
  pp. 1256--1266, 2019.

\bibitem{shi2019furcanext}
Z.~Shi, H.~Lin, L.~Liu, R.~Liu, and J.~Han,
\newblock ``{FurcaNeXt}: End-to-end monaural speech separation with dynamic
  gated dilated temporal convolutional networks,''
\newblock {\em arXiv preprint arXiv:1902.04891}, 2019.

\bibitem{Shi2019deep}
Z.~Shi, H.~Lin, L.~Liu, R.~Liu, J.~Han, and A.~Shi,
\newblock ``Deep attention gated dilated temporal convolutional networks with
  intra-parallel convolutional modules for end-to-end monaural speech
  separation,''
\newblock in {\em Proc. Interspeech}, 2019, pp. 3183--3187.

\bibitem{chen2017cracking}
Z.~Chen, J.~Li, X.~Xiao, T.~Yoshioka, H.~Wang, Z.~Wang, and Y.~Gong,
\newblock ``Cracking the cocktail party problem by multi-beam deep attractor
  network,''
\newblock in {\em Proc. ASRU}, 2017, pp. 437--444.

\bibitem{chen2018efficient}
Z.~Chen, T.~Yoshioka, X.~Xiao, L.~Li, M.~L. Seltzer, and Y.~Gong,
\newblock ``Efficient integration of fixed beamformers and speech separation
  networks for multi-channel far-field speech separation,''
\newblock in {\em Proc. ICASSP}, 2018, pp. 5384--5388.

\bibitem{wang2018multi}
Z.-Q. Wang, J.~Le~Roux, and J.~R. Hershey,
\newblock ``Multi-channel deep clustering: Discriminative spectral and spatial
  embeddings for speaker-independent speech separation,''
\newblock in {\em Proc. ICASSP}, 2018, pp. 1--5.

\bibitem{wang2018combining}
Z.-Q. Wang and D.~Wang,
\newblock ``Combining spectral and spatial features for deep learning based
  blind speaker separation,''
\newblock {\em IEEE/ACM Trans. Audio, Speech, Lang. Process.}, vol. 27, no. 2,
  pp. 457--468, 2018.

\bibitem{chen2019multi}
L.~Chen, M.~Yu, D.~Su, and D.~Yu,
\newblock ``Multi-band {PIT} and model integration for improved multi-channel
  speech separation,''
\newblock in {\em Proc. ICASSP}, 2019, pp. 705--709.

\bibitem{gu2019end}
R.~Gu, J.~Wu, S.-X. Zhang, L.~Chen, Y.~Xu, M.~Yu, D.~Su, Y.~Zou, and D.~Yu,
\newblock ``End-to-end multi-channel speech separation,''
\newblock {\em arXiv preprint arXiv:1905.06286}, 2019.

\bibitem{bahmaninezhad2019comprehensive}
F.~Bahmaninezhad, J.~Wu, R.~Gu, S.-X. Zhang, Y.~Xu, M.~Yu, and D.~Yu,
\newblock ``A comprehensive study of speech separation: Spectrogram vs waveform
  separation,''
\newblock {\em Proc. Interspeech}, Sep 2019.

\bibitem{nakatani2010speech}
T.~Nakatani, T.~Yoshioka, K.~Kinoshita, M.~Miyoshi, and B.-H. Juang,
\newblock ``Speech dereverberation based on variance-normalized delayed linear
  prediction,''
\newblock {\em IEEE/ACM Trans. Audio, Speech, Lang. Process.}, vol. 18, no. 7,
  pp. 1717--1731, 2010.

\bibitem{delfarah2018recurrent}
M.~Delfarah and D.~Wang,
\newblock ``Recurrent neural networks for cochannel speech separation in
  reverberant environments,''
\newblock in {\em Proc. ICASSP}, 2018, pp. 5404--5408.

\bibitem{delfarah2019deep}
M.~Delfarah and D.~Wang,
\newblock ``Deep learning for talker-dependent reverberant speaker separation:
  An empirical study,''
\newblock {\em IEEE/ACM Trans. Audio, Speech, Lang. Process.}, vol. 27, no. 11,
  pp. 1839--1848, 2019.

\bibitem{menne2019analysis}
T.~Menne, I.~Sklyar, R.~Schlüter, and H.~Ney,
\newblock ``Analysis of deep clustering as preprocessing for automatic speech
  recognition of sparsely overlapping speech,''
\newblock {\em Interspeech 2019}, Sep 2019.

\bibitem{lea2016temporal}
C.~Lea, R.~Vidal, A.~Reiter, and G.~D. Hager,
\newblock ``Temporal convolutional networks: A unified approach to action
  segmentation,''
\newblock in {\em European Conference on Computer Vision}. Springer, 2016, pp.
  47--54.

\bibitem{chollet2017xception}
F.~Chollet,
\newblock ``Xception: Deep learning with depthwise separable convolutions,''
\newblock in {\em Proc. IEEE Conf. Computer Vision and Pattern Recognition},
  2017, pp. 1251--1258.

\bibitem{yoshioka2012generalization}
T.~Yoshioka and T.~Nakatani,
\newblock ``Generalization of multi-channel linear prediction methods for blind
  {MIMO} impulse response shortening,''
\newblock {\em IEEE/ACM Trans. Audio, Speech, Lang. Process.}, vol. 20, no. 10,
  pp. 2707--2720, 2012.

\bibitem{allen1979image}
J.~B. Allen and D.~A. Berkley,
\newblock ``Image method for efficiently simulating small-room acoustics,''
\newblock {\em The Journal of the Acoustical Society of America}, vol. 65, no.
  4, pp. 943--950, 1979.

\bibitem{povey2018semi}
D.~Povey, G.~Cheng, Y.~Wang, K.~Li, H.~Xu, M.~Yarmohammadi, and S.~Khudanpur,
\newblock ``Semi-orthogonal low-rank matrix factorization for deep neural
  networks,''
\newblock {\em Proc. Interspeech 2018}, pp. 3743--3747, 2018.

\bibitem{povey2011kaldi}
D.~Povey, A.~Ghoshal, G.~Boulianne, L.~Burget, O.~Glembek, N.~Goel,
  M.~Hannemann, P.~Motlicek, Y.~Qian, P.~Schwarz, et~al.,
\newblock ``The {K}aldi speech recognition toolkit,''
\newblock in {\em Proc. ASRU}, 2011.

\bibitem{seki2018purely}
H.~Seki, T.~Hori, S.~Watanabe, J.~Le~Roux, and J.~R. Hershey,
\newblock ``A purely end-to-end system for multi-speaker speech recognition,''
\newblock {\em Proceedings of the 56th Annual Meeting of the Association for
  Computational Linguistics (ACL)}, pp. 2620--2630, 2018.

\bibitem{chang2019end}
X.~Chang, Y.~Qian, K.~Yu, and S.~Watanabe,
\newblock ``End-to-end monaural multi-speaker {ASR} system without
  pretraining,''
\newblock in {\em Proc. ICASSP}, 2019, pp. 6256--6260.

\bibitem{barker2018fifth}
J.~Barker, S.~Watanabe, E.~Vincent, and J.~Trmal,
\newblock ``The fifth “chime” speech separation and recognition challenge:
  Dataset, task and baselines,''
\newblock {\em Interspeech 2018}, Sep 2018.

\end{thebibliography}

\end{document}